\documentclass[12pt]{article}
\usepackage[cp1251]{inputenc}
\usepackage{amssymb,amsmath,epsfig}
\usepackage{graphicx}
\usepackage{subfig}
\usepackage{cite}
 \allowdisplaybreaks

\begin{document}

\title{\bf Bounce cosmology in $f(\mathcal{R})$ gravity}

\author{M. Ilyas$^1$ \thanks{ilyas\_mia@yahoo.com}
and W. U. Rahman$^2$ \thanks{waheed71rahman@gmail.com}\\
$^1$ Institute of Physics, Gomal University,\\
Dera Ismail Khan, 29220,
Khyber Pakhtunkhwa, Pakistan\\
$^2$ Department of Physics, Abdul Wali khan University Mardan,\\
Mardan, 23200, Pakistan\\}

\date{}

\maketitle

\begin{abstract}
In this paper, we analyze the modified $f(\mathcal{R})$ gravity models in Friedmann--Lema\^{\i}tre--Robertson--Walker (FLRW) background. The actions of bouncing cosmology are studied under consideration of different viable models in $f(\mathcal{R})$ gravity theory that can resolve the difficulty of singularity in standard Big-Bang cosmology. Under different viable models in $f(\mathcal{R})$ gravity theory, the cosmological constraints are plotted in provisions of cosmic-time, then investigated the bounce circumstance. In addition, the red-shift parameter is used to reconstruct the modified gravity, and compile the cosmological parameters that infer accelerated universe expansion. Finally, the situation stability is evaluated with a sound speed feature, which illustrates late-time stability.
\end{abstract}
\maketitle


\section{Introduction}\label{I}

General theory of relativity has explained many specification of universe by the various theories and observational evidence. The observational components of $\Lambda$-Cold dark space model is consistent with all cosmological observations, but it suffers from such differences such as cosmic coincidence or tuning \cite{refa1,refa2,refa3}. Since an unusual drop in the observed energy streams from cosmic radiation base radiation, massive systems, red-shift as well as supernovae Type Ia assessments the rapid expansion of the earth has become apparent \cite{refa4,refa5,refa6}. The reasons behind this curious and riddling phenomena referred to such discoveries as dark energy (DE) (an elusive force). Different methods have been suggested in these directions to change Einstein gravity. So, we should focus our study to other unknown problems. Non-singularity is one of unknown problems in the Big bang cosmology model. On the way to resolve this problem, we require to establish another new approach for the visible universe e.g. the oscillatory universe. Its mean that our universe is collapse of an old universe~\cite{ref1}. Bouncing universe is the new idea, suggested to resolves the singularity problem in big bang cosmology~\cite{ref2,ref3}. Moreover, the bouncing behavior of universe were investigated deeply in the brane cosmology as well as in vector field~\cite{ref4,ref5}. The bounce occurs in the universe when the universe is entered to the big bang era, this phenomenon is the simple interpretation of the bouncing universe (for further literature review, see~\cite{b0,b00,b1,b2,b3,b4,b5,b6,b7,b8,b9,b10,b11,b12,b13,b14,b15,b16,b17,b18,b19,b20,b21,b22,b23,b24,b25,b26,b27,b28,b29,b30,b31,b32,b33,b34,b35,
b36,b37,b38,b39,b40,b41,b42,b43,b44,b45,b46,b47,b48}). The initial phase of the universe transfer to an accelerated expansion phase and at that point the Hubble parameter, $H(t)$, transit from $H(t)<0$ to $H(t)>0$ and also in bounce point, we have $H(t)=0$~\cite{ref6,ref7,ref8,ref9}.\\

Recently, several articles have been studied due to various theories and observational evidence about the splitting of universe while this growing is undergoing to moving an accelerating stage. This study was arise due to the discovery in the type Ia supernova~\cite{ref10} connected with long scale arrangement~\cite{ref11}, and  background with cosmic microwave~\cite{ref12} . We have studied the accelerated growing is appropriate to a mysterious energy (known as the DE) that is approximately $70\%$ of the whole universe. As the negative pressure in universe with perfect fluid and the equation of state (EoS) parameter, $\omega$, which is not greater than $-1$ are known as phantom phase.\\

The energy conditions in nonlocal gravity were studied in Ref \cite{refmia1}, which is helpful in the key role of DE. There are several candidates regarding the structure of DE, some of them are, the scalar field (including tachyon, quintessence, quintom and phantom etc.)~\cite{ref14,ref15,ref16,ref17,ref18,ref19}, cosmological constant, $\Lambda$, ~\cite{ref13}, brane universe models~\cite{ref27,ref28,ref29}, interacting models~\cite{ref24,ref25,ref26}, and holographic models~\cite{ref20,ref21,ref22}.\\

The adjusted gravity of these models has many benefits for many other models, but the most complex numerical approach is avoided by computing science and the current findings of the former accelerating world with dark energies are still present. The simplest modified gravity is $f(\mathcal{R})$ gravity, in which Ricci scalar, $\mathcal{R}$, is replaced by an arbitrary function, $f(\mathcal{R})$, in the Hilbert-Einstein gravitational action~\cite{ref30}.\\

$f(\mathcal{R})$ is actually a family of theories, each one defined by a different arbitrary function. The accelerated expansion and structure formation of the Universe were studied without adding unknown forms of DE or DM. Some models in $f(\mathcal{R})$ gravity may be inspired by corrections arising from the reconciliation of GR with quantum mechanics (e.g. string theories etc.). The modified $f(\mathcal{R})$ gravity was suggested in 1970 and a wide range of phenomena can be produced from this theory by adopting different functions; however, many functional forms can now be ruled out on observational grounds, or because of pathological theoretical problems.\\

The modified $f(\mathcal{R})$ gravity is a best alternative way of instead of the standard model of gravity which is known as a basis of DE (for reviews on modified
$f(\mathcal{R})$ theory of gravity, see~\cite{ref301,ref302,ref303,ref304,ref305,ref306}).\\

In this study, in the big bang theory, we will avoid the starting singularity by bouncing model with the advantage of different models in modified $f(\mathcal{R})$ theory. Furthermore, we will prove that there is one of the speeded phase-shift from a starting contracting-phase to the growing-phase. The problem is well explain by the temporal derivative of the scale-factor which is $\dot{a}(t)$, during contraction decreases $(\dot{a}(t)<0)$  and in the increasing phase  $(\dot{a}(t)>0)$ it increases and at the bounce point it becomes equal to zero $(\dot{a}(t)=0)$. The already mentioned story will be confirmed by the corresponding figures. from the above discussion we say that the dark energy that is reconstructed by red shift parameter is a source of $f(\mathcal{R})$ gravity. Therefore parametrization for $f(\mathcal{R})$ will be found, with the help of motivation accelerated growing of the world can be showed. Finally, the consistency of the model is analyzed such that the subjectivity of adiabatic disorder is assumed to be a thermodynamic framework. Sound speed is used to research the stability of the system as a specific purpose.\\

We attempt to explain  the following the $f(\mathcal{R})$ Gravity to recreate by red-shift factor as an origin of DE. We would to discuss a parameterizations for $f(\mathcal{R})$ gravity than we can define in this inspiration as the expanding of the accelerated universe expansion. Finally, the consistency of the study will be tested such that the University is seen under adiabatic perturbation as either a thermodynamic device. So, the stabilization of the model is studied using a valuable feature called the sound speed.\\

In this paper, we study the modified $f(\mathcal{R})$ gravity and solve the field equations for FLRW metric in the framework of perfect fluid in Sec. \ref{II}. While in Sec. \ref{III}, we analyze the bouncing behavior by the Hubble parameter and scale-factor, by assuming different models in $f(\mathcal{R})$ gravity. Then in Sec. \ref{IV}, the existing red-shift parameter model is reconstructed. Then, effective pressure and effective energy density is redefined in term of red-shift parameter and also the cosmological parameters will also be specified. In Sec. \ref{V}, We investigate and examine the stability of the models. The brief description and conclusion of our results are summarized in last section.

\section{The Modified $f(\cal{R})$ gravity theory}\label{II}

Let us begin by analyzing the implications of the modified $f(\cal{R})$ gravity
\begin{equation}\label{e1}
A_{f(\mathcal{R})}=\int \sqrt{-g}\left[f(\mathcal{R})+2 \kappa^2 L_m\right]d^4x,
\end{equation}
where $f(\cal{R})$ is an arbitrary function of Ricci scalar while $\kappa^2 \equiv 8 \pi G$ is coupling constant, and $L_m$ is Lagrangian matter density. By variational principles, the field equation becomes
\begin{equation}\label{e2}
f_\mathcal{R} R_{\lambda \rho}-\frac{1}{2} f g_{\lambda \rho}-\left[\nabla_\lambda \nabla_\rho- g_{\lambda \rho} \Box\right] f_\mathcal{R} = \kappa^2 T_{\lambda \rho}^{(m)},
\end{equation}

where $\nabla_{\lambda}$, $\Box\equiv \nabla^{\lambda}\nabla_{\lambda}$, $T_{\lambda \rho}^{(m)}$ are the covariant derivative, D'Alembert's operator and the energy-momentum tensor respectively, while $f_\mathcal{R} = \frac{d f}{d \mathcal{R}}$. The
quantity $f_\mathcal{R}$ contains the second order derivatives of the
metric variables and the trace of Eq.(\ref{e2}) gives
\begin{align}\label{p13}
3\Box{f_R}+{R}f_R-2f(R)-{\kappa}^2 T=0,
\end{align}
where $T\equiv T^{(m)\rho}_{\rho}$. The above equation is a second order differential equation in $f_\mathcal{R}$, unlikely the trace of field equation in general theory of relativity is reduces to $\mathcal{R}+\kappa^2 T=0$. This points $f_\mathcal{R}$ as a source of generating scalar
degrees of freedom in $f(\mathcal{R})$ theory.  As $T=0$ this conditions doesn't essentially indicates the constant value (or vanishing) of $\mathcal{R}$, is the dynamics and Eq.(\ref{p13}) is a fruitful mathematical methods to study various interesting and hidden cosmic
arena, e.g. stability,  Newtonian limit and so on. The constraint, constantly Ricci scalar as well as $T_{\alpha\beta}=0$, reduces Eq.(\ref{p13}) as
\begin{align}
\mathcal{R}f_\mathcal{R}-2f(\mathcal{R})=0,
\end{align}
which is known as the Ricci algebraic equation after selecting any viable
formulations of modified $f(\mathcal{R})$ gravity model. By the roots of
the above mention equation, i.e., $\mathcal{R}=\Lambda$ (assume), so Eq.(\ref{p13}) gives
\begin{align}
\mathcal{R}_{\alpha\beta}=\frac{g_{\alpha\beta}\Lambda}{4},
\end{align}

We let FRLW background metric as
\begin{equation}\label{e3}
d{s^2} = a^2d{r^2}+{a^2}[ {d{\theta ^2} + {{r^2}{\sin }^2}\theta d{\phi ^2}}]-d{t^2},
\end{equation}
here $a$ is the scale-factor  that is the function of $t$. Then the Ricci scalar $\mathcal{R}$ for above metric (Eq. \ref{e3}) is
\begin{equation}\label{e4}
\mathcal{R} = 6\dot{H}+ 12 H^2,
\end{equation}
here $H = \frac{\dot{a}}{a}$ is  known as the Hubble parameter while dot represent derivative w.r.t cosmic-time.\\
By solving the field Eq. \eqref{e2} for metric Eq. (\ref{e3}), we get
\begin{subequations}\label{e5}
\begin{eqnarray}
3H^2 f_\mathcal{R} &=& \kappa^2 \rho_m+\frac{1}{2}[\mathcal{R} f_\mathcal{R}-f]-3 H \dot{\mathcal{R}} f_{\mathcal{R}\mathcal{R}},\label{e5-1}\\
-[2 \dot{H}+ 3 H^2]f_\mathcal{R} &=& \kappa^2 p_m-\frac{1}{2}[\mathcal{R} f_\mathcal{R}-f]+\dot{R}^2 f_{\mathcal{R}\mathcal{R}\mathcal{R}}+2 H \dot{\mathcal{R}} f_{\mathcal{R}\mathcal{R}}+\ddot{\mathcal{R}} f_{\mathcal{R}\mathcal{R}},\label{e5-2}
\end{eqnarray}
\end{subequations}
here $\rho_m$, $p_m$ are energy density and pressure respectively. By using conservation equation, $\nabla_\alpha T^{(m){\alpha \beta}} = 0$, along with EoS Parameter, $\omega_m= \frac{p_m}{\rho_m}$, one find
\begin{equation}\label{e6}
\dot{\rho}_m + 3  \rho_m(\omega_m + 1)H = 0,
 \end{equation}
the above equation yields the solution as
\begin{equation}\label{e7}
\rho_m = \rho_{m_0} a^{-3(\omega_m + 1)}.
\end{equation}

Furthermore, by using the standard Friedmann equations to compare the present approach
$$\rho_{eff} =  \frac{3}{\kappa^2} H^2$$ and $$p_{eff} = -\frac{1}{\kappa^2} (3 H^2+2 \dot{H}),$$
we can rewrite Eqs. \eqref{e5} as
\begin{subequations}\label{e8}
\begin{eqnarray}
\rho^{eff} = \rho_m +\rho_{f(\mathcal{R})}  = \kappa^{-2} \left[\kappa^2 \rho_m+3 H^2 (1-f_\mathcal{R})-\frac{1}{2}(f-\mathcal{R} f_\mathcal{R})-3 H \dot{\mathcal{R}} f_{\mathcal{R}\mathcal{R}}\right],\label{e8-1}\\
\begin{aligned}
p^{eff} = p_m + p_{f(\mathcal{R})} = \kappa^{-2} \bigg[\kappa^2 p_m-(3 H^2+2 \dot{H})(1-f_\mathcal{R})+&\frac{1}{2}(f-\mathcal{R} f_\mathcal{R})+\dot{\mathcal{R}}^2 f_{\mathcal{R}\mathcal{R}\mathcal{R}}\label{e8-2}\\
&+2 H \dot{\mathcal{R}} f_{\mathcal{R}\mathcal{R}}+\ddot{\mathcal{R}} f_{\mathcal{R}\mathcal{R}}\bigg],
\end{aligned}
\end{eqnarray}
\end{subequations}
here $\rho^{eff}$, $p^{eff}$ are effective energy density and pressure, while

\begin{subequations}\label{e9}
\begin{eqnarray}
&\rho_{f(\mathcal{R})} = \kappa^{-2} \left[3 H^2 (1-f_\mathcal{R})-\frac{1}{2}(f-\mathcal{R} f_\mathcal{R})-3 H \dot{\mathcal{R}} f_{\mathcal{R}\mathcal{R}}\right],\label{e9-1}\\
&p_{f(\mathcal{R})} = \kappa^{-2} \left[-(2 \dot{H}+3 H^2)(1-f_\mathcal{R})+\frac{1}{2}(f-\mathcal{R} f_\mathcal{R})+\dot{\mathcal{R}}^2 f_{\mathcal{R}\mathcal{R}\mathcal{R}}+2 H \dot{\mathcal{R}} f_{\mathcal{R}\mathcal{R}}+\ddot{\mathcal{R}} f_{\mathcal{R}\mathcal{R}}\right].\label{e9-2}
\end{eqnarray}
\end{subequations}

In the case of effective terms, the conservation equation can be rewritten with the help of using Eqs. \eqref{e8} as
\begin{equation}\label{e10}
\dot{\rho}^{eff} + 3\rho^{eff} H (1 + \omega^{eff}) = 0,
 \end{equation}
Here
 \begin{equation}\label{e10-1}
\omega^{eff} = \frac{p^{eff}}{\rho^{eff}} = -\left(1+\frac{2 \dot{H}}{3 H^2}\right),
\end{equation}
is the parameter of effective EoS .

\section{Reconstruction Method for Power-law model}
By using the method in Refs \cite{rec1,rec2}, we reconstruct the modified gravitational models with the help of introducing proper functions $P(t)$ and $Q(t)$ of a scalar field $t$, which is interpreted as the cosmic time, the action in Eq. (\ref{e1}) with the absence of matter is written as
\begin{equation}
I = \int {d{x^4}\sqrt { - g} \left[ {P(t) R+ Q(t) + {L_m}} \right]}
\end{equation}
then by solving this equation with eq. (\ref{e5}) we get

\begin{equation}
Q(t) =  - 6{H^2}P'(t) - 6P(t){H^2}(t)
\end{equation}
while the second equation becomes
\begin{equation}
P''(t) - H(t)P'(t) + 2P(t)H'(t) = 0
\end{equation}

\begin{figure} \centering
\epsfig{file=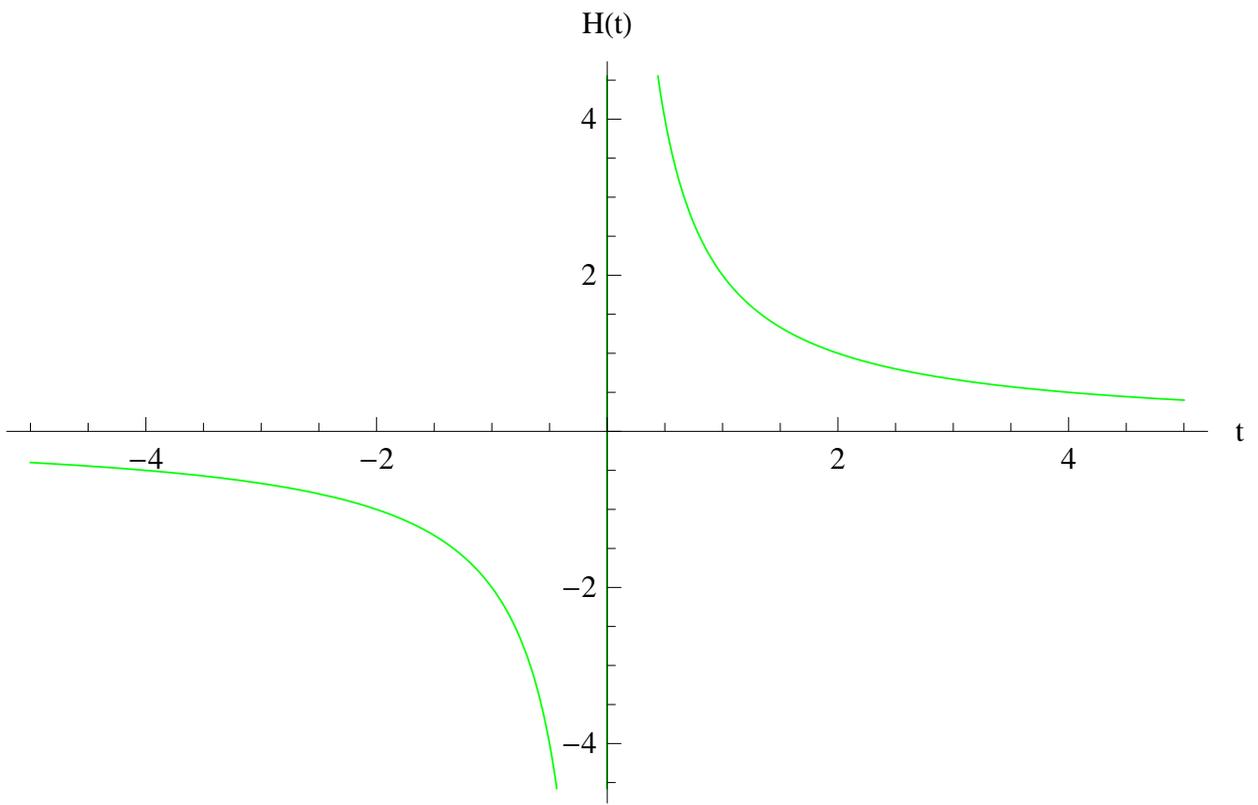,width=1\linewidth}
\caption{Hubble parameter for Power Law}\label{h3}
\end{figure}

We solve the above differential equation for the power-law scale factor and then we use the solution to write down the general form of $f(R)$ as follow
\begin{equation}\label{rec1}
f(R)=P(R)R + Q(R)
\end{equation}
The scale factor for power law model is given by
\begin{equation}
a(t) = \beta {t^{\left( {2n} \right)}}
\end{equation}
Where $\beta$ is constant and $n$ is an integer. The behavior of this type of bouncing is shown in fig.(\ref{h3})

By using this scale factor and solving for $P(t)$ and $Q(t)$, we get.

\begin{align}
P(t) &= {t^{n - \frac{1}{2}\sqrt {4n\left( {n + 5} \right) + 1}  + \frac{1}{2}}}\left( {{c_1}{t^{\sqrt {4n\left( {n + 5} \right) + 1} }} + {c_2}} \right),\\
Q(t) &= 6n{t^{\frac{1}{2}\left( {2n - \sqrt {4n\left( {n + 5} \right) + 1}  - 3} \right)}}\left[ {{c_2}\left( { - 6n + \sqrt {4n\left( {n + 5} \right) + 1}  - 1} \right)} \right.\\
 &- {c_1}\left( {6n + \sqrt {4n\left( {n + 5} \right) + 1}  + 1} \right)\left. {{t^{\sqrt {4n\left( {n + 5} \right) + 1} }}} \right]
\end{align}

The cosmic time in term of ricci scalar is written as

\begin{equation}
{t_ \pm } = \left[ {\frac{{ \pm 2\sqrt 3 \sqrt {n\left( { - 1 + 4n} \right)} }}{{\sqrt R }}} \right]
\end{equation}
By putting these equations in Eq. (\ref{rec1}), we get
\begin{align}\label{mod3}
f(R) &= \frac{1}{{4n - 1}}{2^{\frac{1}{2}\left( {2n - \sqrt {4n\left( {n + 5} \right) + 1}  - 1} \right)}}{3^{\frac{1}{4}\left( {2n - \sqrt {4n\left( {n + 5} \right) + 1}  + 1} \right)}}\\
&R{[ {\frac{{\sqrt {n\left( {4n - 1} \right)} }}{{\sqrt R }}} ]^{n - \frac{1}{2}\sqrt {4n\left( {n + 5} \right) + 1}  + \frac{1}{2}}}\left[ {{c_2}(2n} \right. + \sqrt {4n\left( {n + 5} \right) + 1}  - 3)\\
&- {c_1}{2^{\sqrt {4n\left( {n + 5} \right) + 1} }}{3^{\frac{1}{2}\sqrt {4n\left( {n + 5} \right) + 1} }}\left\{ { - 2n} \right. + \left. {\sqrt {4n\left( {n + 5} \right) + 1}  + 3} \right\}\\
&\left. {{{[ {\frac{{\sqrt {n\left( {4n - 1} \right)} }}{{\sqrt R }}}]}^{\sqrt {4n\left( {n + 5} \right) + 1} }}} \right]
\end{align}

Here, $c_1,c_2$ are constants. We can see that this is valid for every $n$ except $0\leq n \leq-5$ and for $n=1$ we get

\begin{equation}
{f_1}(R) = \frac{1}{9}{c_2}{R^{3/2}} - \frac{{{c_1}}}{R}
\end{equation}

\section{The Bouncing behavior in modified $f(\mathcal{R})$ gravity}\label{III}

Here, we will investigate the rebound conditions in modified $f(\mathcal{R})$ gravity. As an evolving world emerges, the world transitions fluctuations into an expanding stage from an earlier contracting period. This stage transition leads to a non-singular outcome in the Big Bang cosmological norm \cite{ref6, ref7, ref8, ref9}.

 Consequently, the Hubble parameter moves for a good bounce from $H(t) < 0$ to $ H(t) > 0$ and in bounce point $H(t) = 0$. We may also claim the dilemma of bouncing World as regards the scale-factor, i.e. that we have a decline in scale-factor even during contracting stage as $\dot{a}(t) < 0$, and we have a development in the expanding process $\dot{a}(t) > 0$,and at the point of bounce $\dot{a}(t) =  0 $ and this point around $\ddot{a}(t)  >  0$.

Subsequently the Hubble parameter moves from $H(t)<0$ to $H(t)>0$, since its derivative must be greater than zero in the bounce stage as regards time evolution for its bouncing world
\begin{equation}\label{e11}
\dot{H}_{bounce}=-\frac{\kappa^2}{2}(1+\omega^{eff}) \rho^{eff} > 0,
\end{equation}
We deduce the condition $\rho^{eff}>0$ with $\omega^{eff}<-1$.

Now within the bounce stage we can get the bounce status for the model. We would have in this situation $\dot{H}_{bounce}>0$ with $H_{bounce}=0$, such that the bouncing stage \eqref{e11} in the bounce stage by Eqs. \eqref{e8} be assumed as
\begin{equation}\label{e12}
\dot{H}_{bounce}=(2 f_\mathcal{R})^{-1}\left[-\kappa^2(\rho_m+p_m)+\dot{\mathcal{R}}^2 f_{\mathcal{R}\mathcal{R}\mathcal{R}}+\ddot{\mathcal{R}} f_{\mathcal{R}\mathcal{R}}\right]>0.
\end{equation}
In the following, we will investigate bouncing behavior with the advantage of particular choices in $f (\mathcal{R})$ models

\begin{itemize}
  \item Model 1\\
  We consider the model with quadratic corrections firstly proposed by Starobinsky \cite{refmod1}
\begin{equation}\label{model1}
f(\mathcal{R})=\mathcal{R} + \alpha \mathcal{R}^2
\end{equation}
\\
where $\alpha$ is constants.
  \item Model 2\\
we take the Ricci scalar exponential corrections to GR \cite{refmod2}\\
\begin{equation}\label{model2}
f(\mathcal{R})=\mathcal{R} + \alpha  \mathcal{R}  \left(e^{(-\frac{\mathcal{R}}{\gamma})} - 1\right)
\end{equation}
\\
where $\alpha$ and $\gamma$ are constants.
  \item Model 3\\
One of the cubic correction to GR \cite{refmod3}\\
\begin{equation}\label{model3}
f(\mathcal{R})=\mathcal{R} + \alpha \mathcal{R}^2 \left(1 + \gamma \mathcal{R}\right)
\end{equation}
\\
where  $\gamma$ and $\alpha$ are arbitrary constants and if  $\gamma >> R$, Therefore the quadratic model is suitable  model. This one. the case in doubt $\alpha \mathcal{R}^2 \left(1 + \gamma \mathcal{R}\right) <<R$ can be employed to maximise the influence of cubic terms on quadratic terms.
  \item Model 4\\
\begin{equation}\label{model4}
f(\mathcal{R})=\mathcal{R} + \alpha \mathcal{R}^2 \left(1 + \gamma  Log[\frac{\mathcal{R}}{\mu^2}]\right)
\end{equation}
This model contain the logarithm term which describe the universe evolution in the absence of DE \cite{refmod4}.
\end{itemize}

\begin{figure}
\begin{center}
\epsfig{file=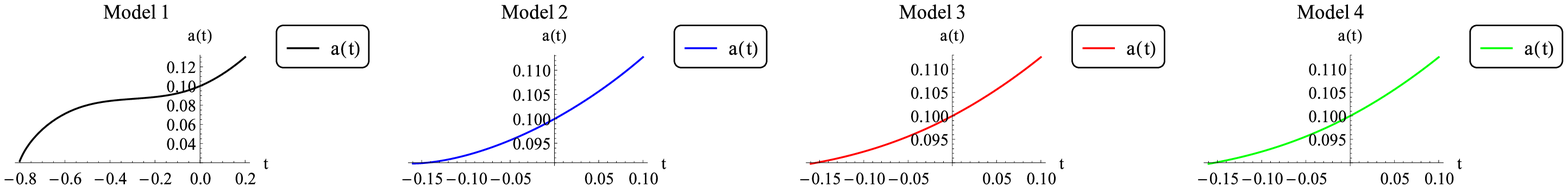,width=1\linewidth}
\caption{The scale-factor, $a(t)$, in terms of cosmic-time under different viable models.}\label{fig1}
\end{center}
\end{figure}

\begin{figure}
\begin{center}
\epsfig{file=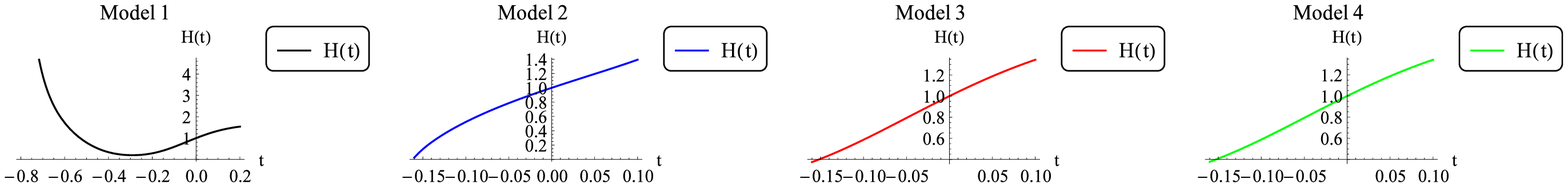,width=1\linewidth}
\caption{The Hubble parameter, $H(t)$, in terms of cosmic-time under different viable models.}\label{fig2}
\end{center}
\end{figure}

\begin{figure}
\begin{center}
\epsfig{file=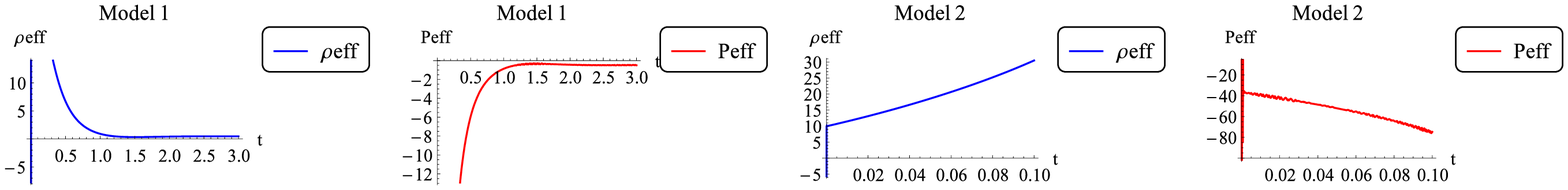,width=1\linewidth}
\caption{The energy density, $\rho^{eff}$, and pressure, $p^{eff}$, in terms of cosmic-time under different viable models.}\label{fig3}
\end{center}
\end{figure}

\begin{figure}
\begin{center}
\epsfig{file=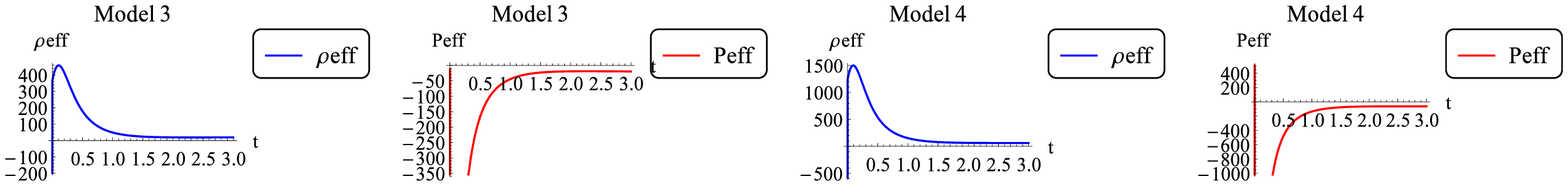,width=1\linewidth}
\caption{The energy density, $\rho^{eff}$, and pressure, $p^{eff}$, in terms of cosmic-time under different viable models.}\label{fig3}
\end{center}
\end{figure}

\begin{figure}
\begin{center}
\epsfig{file=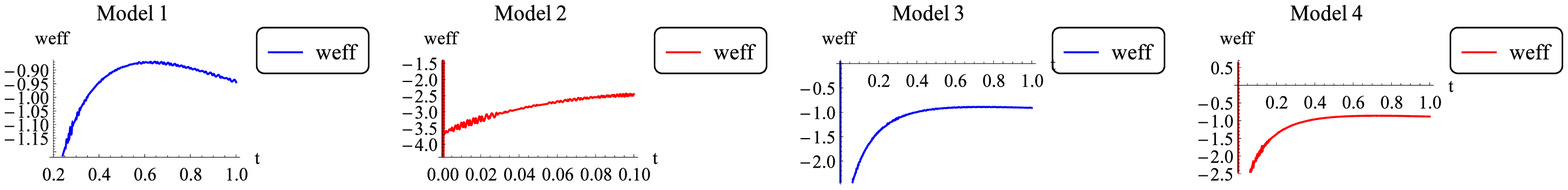,width=1\linewidth}
\caption{The EoS parameter, $\omega^{eff}$, in terms of cosmic-time under different viable models.}\label{fig5}
\end{center}
\end{figure}

Moreover, by inserting models (\ref{model1}),(\ref{model2}),(\ref{model3}) and (\ref{model4}) into Eqs. \eqref{e8} separately, and solving numerically. By plotting the numerical solutions, the cosmological parameters can easily be drawn in terms of cosmic-time as shown in Figs. \ref{fig1}-\ref{fig2}.

The bouncing behavior under consideration of four different viable models are:\\

In case of model 1, we can detect the bouncing action in the first part of Fig. \ref{fig1}. The Hubble parameter moves from the point of bounce ($H(t \sim -0.818)=0$) as $H(t)<0$ to $H(t)>0$, while the lowest scale-factor is or $\dot{a}(t \sim -0.818)=0$.\\

In Fig. \ref{fig1}, we can detect the bouncing action like $H(t \sim -0.162019)=0$ as $H(t)<0$ to $H(t)>0$, and in other hand side, we see that the minimal element for the scale is or $\dot{a}(t \sim -0.162019)=0$ for the case of model 2.\\

Similarly, we realize that the bouncing behavior can be seen in the figure. \ref{fig2}, in which we see that from the bounce level, the Hubble parameter moves through $H(t \sim -0.354734)=0$ as $H(t)<0$ to $H(t)>0$, and In another hand, we can see how the minimal element for the scale or $\dot{a}(t \sim -0.354734)=0$ for the case of model 3.\\

Furthermore, we believe it is possible to detect the bouncing behavior in Fig. \ref{fig2}, in which the Hubble parameter moves $H(t \sim -0.281699)=0$ as $H(t)<0$ to $H(t)>0$, and in other hand, we can see how the minimal element for the scale or $\dot{a}(t \sim -0.281699)=0$ for the case of model 4.

The Fig. \ref{fig3} and \ref{fig5} show that $\rho_{eff} > 0$ and $p_{eff} < 0$, These illustrate an accelerating universe. We may also note EoS variations with reference to galactic time in fig. \ref{fig6}, and one indicates that the problem relates to late-time observational data crossing from over phantom-divide-line \cite{ref33, ref34}. \\
We will rebuild the above model using the definition of red-shift during the next part.

\section{Reconstruction by red-shift parameter}\label{IV}

In this part, we will analyze the red-shift parameter model. In view of the scale-factor, one will be added as $z+1=\frac{a_0}{a(t)}$, here $a_0$ in the value in the current time. Further, we assume the dimensionless parameters $r(z)=\frac{H(z)^2}{H_0^2}$ where $H_0=71\pm3\,km\,s^{-1}\,Mpc^{-1}$, is the Hubble value in the present time (universe today). From the above, one can relate cosmic-time to red-shift, the differential form appears
\begin{equation}\label{e19}
\frac{d}{dt}=\frac{da}{dt}\frac{dz}{da}\frac{d}{dz}=-H(z + 1) \frac{d}{dz},
\end{equation}
Furthermore, the Ricci scalar (Eq. \eqref{e4}) and the effective energy density and pressure (Eq. \eqref{e8}) can be rewritten in terms of $z$ as
\begin{equation}\label{e19}
\mathcal{R}=12H_0^2 r-3 (z+1)H_0^2 r',
\end{equation}
\begin{eqnarray}\label{rhototz}
\begin{aligned}
\rho^{eff}=\kappa^{-2} \Big[\kappa^2 \rho_m+3 H_0^2 r(f_\mathcal{R}(z)+1)- \frac{3}{2}H_0^2(z+1) r' f_\mathcal{R}(z)+ 3 H_0^2(z+1) r &f'_\mathcal{R}(z)\\
&-\frac{1}{2} f(z) \Big],
\end{aligned}
\end{eqnarray}
\begin{eqnarray}\label{ptotz}
\begin{aligned}
p^{eff}=\kappa^{-2} \Big[\kappa^2 p_m+\frac{1}{2} & H_0^2 (z+1) r' (f_\mathcal{R}(z)+2) - 3 H_0^2 r (f_\mathcal{R}(z)+1) + \frac{1}{2}f(z)\\
&+H_0^2 (z+1)^2 r f''_\mathcal{R}(z) + \frac{1}{2}H_0^2 (z+1)^2 r' f'_\mathcal{R}(z) - H_0^2 (z+1) r f'_\mathcal{R}(z) \Big],
\end{aligned}
\end{eqnarray}
here derivative is denoted by the prime with regard to $z$. The energy density of matter is identified as
\begin{equation}
 \rho_m =  \rho_{m_0} \left(\frac{a_0}{1+z}\right)^{-3(\omega_m+1)}.
\end{equation}

Currently, in order to recreate the model as a source of DE, we need to add the $r(z)$ feature to be equipped with supernova descriptive statistics \cite{ref35, ref36}. In form of red-shift, one of the suitable choice \cite{ref37, ref38}, written as following
\begin{equation}\label{rz}
  r(z)=C_{{0}}+C_{{1}} \left( z+1 \right) +C_{{2}} \left( z+1 \right) ^{2}+\Omega_{{m_{{0}}}} \left( z+1 \right) ^{3},
\end{equation}
where $C_0=1 - C_1 - C_2 - \Omega_{m_0}$. It should be remembered that the above parametrization corresponds to $\Lambda$CDM model for $C_1 = C_2 = 0$ along with $C_0 =1 - \Omega_{m_0}$.  The criteria for the right fit are as good as $\Omega_{m_0} = 0.3$, $C_1=-4.16 \pm 2.53$ and $C_2 = 1.67 \pm 1.03$ \cite{ref33}. It is to be noted that the freely parameters of the model play a crucial role in this work.\\

\begin{figure}
\begin{center}
\epsfig{file=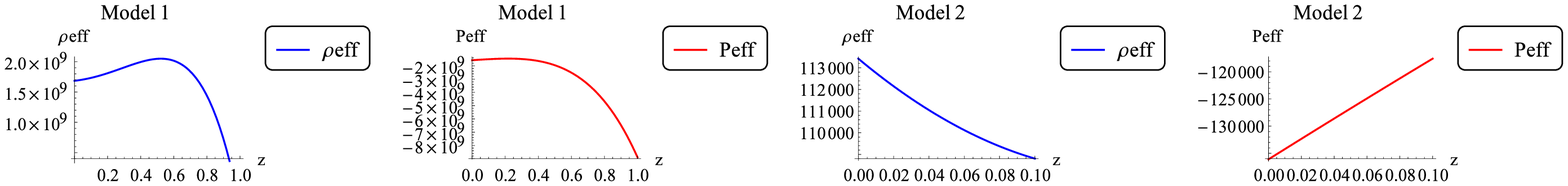,width=1\linewidth}
\caption{The energy density $\rho^{eff}$ and pressure $p^{eff}$ in terms of red-shift under different viable models.}\label{fig6}
\end{center}
\end{figure}

\begin{figure}
\begin{center}
\epsfig{file=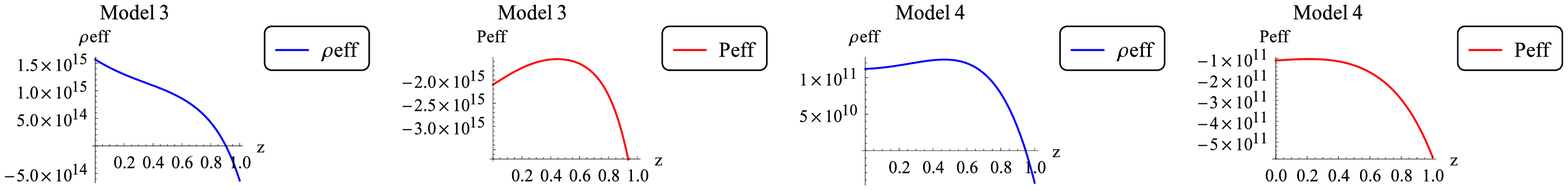,width=1\linewidth}
\caption{The energy density $\rho^{eff}$ and pressure $p^{eff}$ in terms of red-shift under different viable models.}\label{fig7}
\end{center}
\end{figure}

\begin{figure}[h]
\begin{center}
\epsfig{file=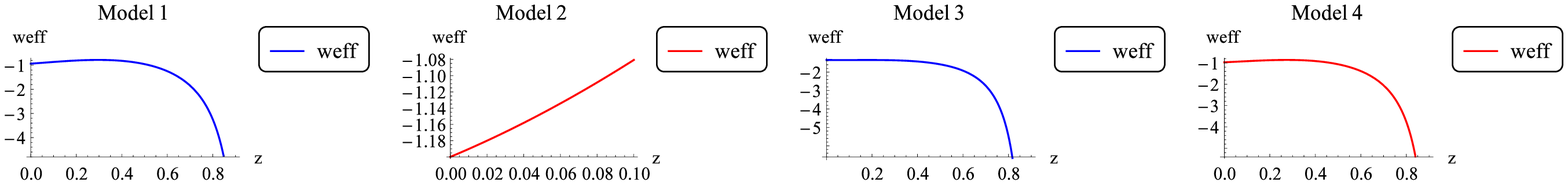,width=1\linewidth}
\caption{The  parameter of EoS is $\omega^{eff}$ in terms of red-shift under different viable models.}\label{fig9}
\end{center}
\end{figure}

By substituting, Eqs. \eqref{rz} into Eqs. \eqref{rhototz} and \eqref{ptotz}, we get the cosmological parameters in terms of red-shift with the results shown in Fig. \ref{fig6} and \ref{fig7}. The variance of the successful EoS tells us that its value is around $\sim-1$ in late time ($z=0$), as shown in Fig. \ref{fig9}, and $\omega^{eff}<-1$ which confirms as an accelerated growing of the universe. The outcome achieved verifies the performance of Refs. \cite{ref39, ref40}.

\section{Stability Analysis}\label{V}
We will address the stability of the modified $f(\mathcal{R})$ gravity. As the ideal fluid has been filled by the Cosmos, we can accept it as a thermodynamic device. We will use the sum of sound-velocity for the ideal fluid device to this end. As we observe, a useful function such as $C_s^2=\frac{dp^{eff}}{d\rho^{eff}}$ induces sound-velocity, in which $p^{eff}$ and $\rho^{eff}$ are the Universe's efficient power density and effective strain. In a thermodynamic system, because the sound-velocity $C_s^2$ is positive, the stability state thus occurs when the parameter $C_s^2$ becomes greater than zero. We observe that a thermodynamic system can be represented by quantities of effective power density, entropy and  effective pressure, with adiabatic and non-adiabatic disruptions.

We now assume the related scheme to be $p^{eff}=p^{eff}(S,\rho^{eff})$, And disturbing ourselves with regard to the successful burden that we have
\begin{equation}\label{effpp}
  \delta p_{eff}=\left(\frac{\partial p_{eff}}{\partial S}\right)_{\rho^{eff}} \delta S+\left(\frac{\partial p^{eff}}{\partial \rho^{eff}}\right)_{S} \delta \rho^{eff} = \left(\frac{\partial p_{eff}}{\partial S}\right)_{\rho^{eff}} \delta S+ C_s^2 \, \delta \rho^{eff},
\end{equation}
In which the first phrase in the cosmological dilemma is compared to a non-adiabatic method, the second term is compared to the adiabatic process. Because adiabatic disruption is considered in cosmology, so for the cosmological process, the variety of entropy appears zero, i.e. $\delta S = 0$. Therefore, we are continuing our study that only involves adiabatic processes.

\begin{figure}
\begin{center}
\epsfig{file=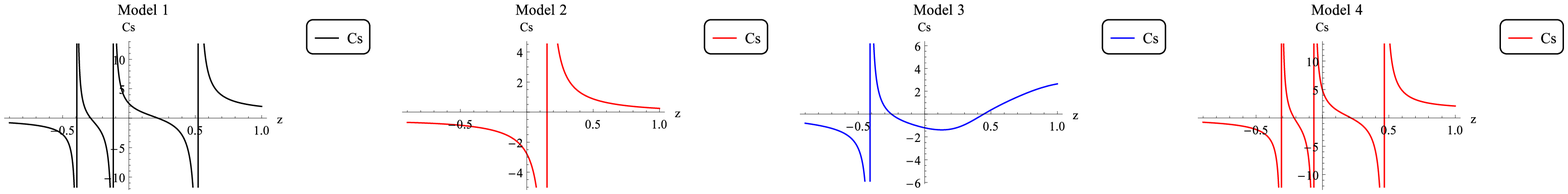,width=1\linewidth}
\caption{The speed of sound $C_s^{2}$ in terms of red-shift under different viable models.}\label{fig10}
\end{center}
\end{figure}

Now, to get the $C_s^2$ feature, to separate the Eqs. With regard to red-shift, \eqref{rhototz} and \eqref{ptotz}, we get $C_s^2$ in terms of red-shift. In that case, by numerical estimation, as seen in Fig. \ref{fig10}, we map the  speed of sound function in the process of red-shift. Hence, the Fig. \ref{fig10} tells us that there is a busy time stability, since the $C_s^2$ value is positive for the $z=0$ event.\\
In case of Model 1, the stability condition satisfied in late-time e.g. at $z=0$, $C_s^2=2.49$ and also $C_s^2>0$ for $z>0.5$, while in case of model 2 and 3, the condition is fulfilled at $z>0.5$ gives $C_s^2>0$. Furthermore, in case of Model 4, the stability condition satisfied in late-time e.g. at $z=0$, $C_s^2=4.9$ and also $C_s^2>0$ for $z>0.5$.

\section{Summary}\label{VI}

We have discussed multiple feasible models of modified $f(\mathcal{R})$ gravity in the FLRW metric. The modified Friedmann equations were obtained by solving the field equations with the perfect fluid. Then, by splitting the two $\rho^{eff}$ and $p^{eff}$ functions for the adjusted gravity, we obtained the efficient EoS. In what follows, in terms of cosmic-time, we studied bouncing behavior for four different viable models in $f(\mathcal{R})$ gravity for the scenario and acquired bouncing state at bounce stage and illustrated the related cosmological parameters. We then rebuilt the model with a red-shift component, and we are using the $r(z)$ function parametrization, and the cosmological parameters were written in terms of red-shift $z$, particularly the Friedmann equations and the effective EoS. The behavior of effective energy density, pressure and EoS in red-shift were studied which tell us that the $\rho^{eff}$ and $p^{eff}$ varieties are positive and negative, respectively. On the other side, the variance of $\omega^{eff}$ tell us that the EoS crosses the phantom step. This outcome supports the gradual expansion of the Universe and thus correlates to observational results. \cite{ref39, ref40}. We observed that in the subsequent graphs, the free parameters play a crucial role in encouraging these choices based on positivity, effective energy density and negative effective pressure, as well as crossing the $\omega^{eff}$ over the phantom divide axis. Finally, we tried to quantify speed sound in terms of red-shift in order to test the stability of the situation, so we plotted the $Cs^2$ with regard to red-shift. Hence, the Fig. \ref{fig10} has shown us that late time stability occurs since the $Cs^2$ function is greater than zero in the actual time. By considering the four different models in modified $f(\mathcal{R})$ gravity, the obtained results also verifies the results of Ref \cite{ref41}.

\end{document}